\documentclass[prd,byrevtex
,preprint
,showkeys
,showpacs]{revtex4}
\usepackage{amsmath,amsfonts}
\usepackage[dvipdfm]{graphicx}
\usepackage{float}
\newcommand{\bs}[1]{\boldsymbol{#1}}

\newcommand{\al}{\alpha}
\newcommand{\del}{\delta}
\newcommand{\ep}{\epsilon}
\begin{document}
\title{Entanglement of Coarse Grained Quantum Field in the Expanding Universe}
\author{Yasusada Nambu}
\email{nambu@gravity.phys.nagoya-u.ac.jp}
\author{Yuji Ohsumi}
\email{osumi@gravity.phys.nagoya-u.ac.jp}
\affiliation{Department of Physics, Graduate School of Science, Nagoya 
University, Chikusa, Nagoya 464-8602, Japan}

\date{December 7, 2009} 
\begin{abstract}
  We investigate the entanglement of a quantum field in the expanding
  universe. By introducing a bipartite system using a coarse-grained
  scalar field, we apply the separability criterion based on the
  partial transpose operation and numerically calculate the bipartite
  entanglement between separate spatial regions. We find that the
  initial entangled state becomes separable or disentangled after the
  spatial separation of two points exceed the Hubble horizon. This
  provides the necessary condition for the appearance of classicality
  of the quantum fluctuation. We also investigate the condition of
  classicality that the quantum field can be treated as the classical
  stochastic variables.
\end{abstract}
\maketitle

\section{Introduction}

Inflation provides the mechanism to generate the seed fluctuation
which leads to the formation of the large scale structure in our
present Universe. During the accelerated expansion stage of the
inflationary universe, short wavelength quantum fluctuations of the
inflaton field are generated by particle creations and then they
becomes longwavelength fluctuations larger than the Hubble horizon by
the cosmic expansion. The generated longwavelength fluctuations are
considered as the classical fluctuations responsible for the origin of
structure in our Universe. The important question is how such quantum
fluctuations change to classical fluctuations; quantum fluctuations
must acquire the classical stochastic nature and transfer to
classical density perturbations which lead to the gravitational
instability to form the nonlinear structure in the Universe. We must
explain what kind of mechanism causes such a quantum to classical
transition of primordial fluctuations\cite{GuthAH:PRD32:1985,SakagamiM:PTP79:1988,BrandenbergerRH:MPLA5:1990,NambuY:PL276B:1992,AlbrechtA:PRD50:1994,PolarskiD:CQG13:1996,LesgourguesJ:NPB497:1997,KieferC:CQG15:1998}.

In this paper, we aim to investigate this problem from the view-point
of the quantum correlation, entanglement. The entanglement is the
specific nature of the quantum system. When we calculate a correlation
function of observables, we have a possibility that the correlation
function cannot be reproduced using a classical probability
distribution function if the system is entangled and the classical
locality is violated~\cite{EinsteinA:PR47:1935,BellJS:P1:1964}. Thus,
we cannot regard the quantum fluctuations as the classical stochastic
fluctuations as long as the system is entangled. If the quantum
fluctuation becomes classical, the entanglement must be lost.  We
consider the entanglement of the quantum field between two spatially
separated regions in the expanding universe and investigate how the
quantum fluctuation acquires the classical nature during inflation.
For the quantum field to behave as the classical stochastic field, it
is necessary to lose the quantum correlation and the classical
distribution function must appear.  The entanglement property for the
general $N$-partite system is complicated and we do not have a general
tool to treat such a system.  For a bipartite system, however, we have
the necessary and sufficient conditions for the existence of quantum
correlation (entanglement)
\cite{PeresA:PRL77:1996,HorodeckiP:PLA232:1997,SimonR:PRL84:2000,DuanL:PRL84:2000}
and apply this criterion to our problem.

In our previous work~\cite{NambuY:PRD78:2008}, we investigated the
behavior of the entanglement of the quantum field using a lattice
model of a scalar field. We considered the entanglement between
spatially separated two blocks and found that the bipartite
entanglement between these two blocks is lost after their separation
exceeds the horizon length. We also discussed that the disappearance
of the entanglement yields only the necessary condition for the
quantum fluctuation to be classical. For the establishment of the
classicality of the quantum fluctuation, the existence of the
classical distribution function which reproduces any correlation
function of the quantum field is necessary. We presented this
condition of the classicality in terms of the symplectic eigenvalue of
the covariance matrix ~\cite{NambuY:PRD78:2008}.  In this paper, we
prepare a bipartite system for the scalar field using the coarse
graining of the quantum field.  As the coarse graining, we introduce
the infrared and the ultraviolet cut off of the Fourier expansion of
the scalar field. This coarse graining formally corresponds to the
stochastic approach to inflation~\cite{StarobinskiA:1986} which
derives the quantum dynamics of the long wavelength mode of the scalar
field as the classical Langevin equation.  We investigate the
entanglement and the condition of the classicality for the
coarse-grained scalar field. Especially, we concentrate on the effect
of the expansion rate of the Universe and the effect of the mass of
the scalar field on the entanglement.  We further investigate the
condition of the classicality and look for the condition of the
appearance of the classical stochastic nature.


This paper is organized as follows: In Sec.~II, we first introduce the
concept of the bipartite entanglement and the condition of the
classicality. Then, we define the bipartite system for the scalar
field via coarse-graining. In Sec.~III, we investigate the
entanglement of the scalar field in the Minkowski spacetime. In
Sec.~IV, we consider the entanglement of the scalar field in the
expanding universe and investigate the effect of the expansion rate
and the mass on the entanglement. We further derive the condition for
the classicality. Section V is devoted to the summary and
conclusion. We use units in which $c=\hbar=8\pi G=1$ throughout the
paper.

\section{Formalism}
\subsection{Bipartite entanglement and condition of classicality}
In this paper, we focus on a bipartite system composed of two Gaussian
modes. A quantum state $\hat\rho$ of the bipartite system is defined
to be separable if and only if $\hat\rho$ can be expressed in the
following direct product form
\begin{equation}
  \hat\rho=\sum_jp_j\hat\rho_{jA}\otimes\hat\rho_{jB},\quad\sum_jp_j=1,\quad
  p_j\ge 0.
\end{equation}
where $\hat\rho_{jA}$ and $\hat\rho_{jB}$ are density operators of the
modes of subsystem A and B. If the state of the system cannot be
expressed in this form, the quantum state of the system is
entangled. If the state is entangled, the observables associated to
parties A and B are correlated and their correlations cannot be
reproduced with purely classical means. This leads to the phenomena
peculiar to the quantum mechanics such as the EPR
correlation~\cite{EinsteinA:PR47:1935} and the violation of Bell's
inequality~\cite{BellJS:P1:1964}.

For a bipartite Gaussian state with two modes, we have the necessary and
sufficient conditions for the separability and we can judge whether
the system is entangled or not using these criteria. We adopt in this
paper a criterion based on the partial transpose operation for a
bipartite
system\cite{HorodeckiP:PLA232:1997,SimonR:PRL84:2000,DuanL:PRL84:2000}.
The canonical variables and the commutation relations for the
bipartite system with two modes are expressed as
\begin{equation}
  \label{eq:2party}
  \bs{\hat\xi}=\begin{pmatrix}\hat q_A\\ \hat p_A\\ \hat q_B\\ \hat p_B\end{pmatrix},\qquad
  [\hat\xi_j,\hat\xi_k]=i\Omega_{jk},\quad j,k=1,2,3,4
  \end{equation}
where
\begin{equation}
\bs{\Omega}=\begin{pmatrix}\bs{J} & \bs{0}\\\bs{0}
  &\bs{J}\end{pmatrix},\qquad \bs{J}=
  \begin{pmatrix} 0 & 1 \\ -1 & 0\end{pmatrix}.
\end{equation}
The Gaussian state  is completely specified by the following covariance matrix
\begin{equation}
  \label{eq:covariance}
  V_{jk}=\frac{1}{2}\left\langle\hat\xi_j\hat\xi_k+\hat\xi_k\hat\xi_j\right\rangle
  =\frac{1}{2}\mathrm{Tr}\left(\left(\hat\xi_j\hat\xi_k
      +\hat\xi_k\hat\xi_j\right)\hat\rho\right)
\end{equation}
where we assume the state with $\langle\hat\xi_j\rangle=0$. For a
physical state, the density matrix must be non-negative and the
corresponding covariance matrix must satisfy the inequality~\cite{SimonR:PRL84:2000}
\begin{equation}
  \bs{V}+\frac{i}{2}\bs{\Omega}\ge 0 \label{eq:phys-cond}
\end{equation}
which is the generalization of the uncertainty relation between two
canonically conjugate variables. The separability of the bipartite
Gaussian state is expressed in terms of the partially transposed
covariance matrix $\tilde{\bs{V}}$ obtained by reversing the sign of
party B's momentum. The necessary and sufficient condition of the
separability is given by the
inequality~\cite{SimonR:PRL84:2000,DuanL:PRL84:2000}
\begin{equation}
  \tilde{\bs{V}}+\frac{i}{2}\bs{\Omega}\ge 0, \label{eq:sepa-cond}
\end{equation}
which represents the physical condition for the partially transposed
state. 

The covariance matrix can be diagonalized by an appropriate symplectic
transformation
$\bs{S}\in\mathrm{Sp}(4,\bs{R}),~\bs{S}\bs{\Omega}\bs{S}^T=\bs{\Omega}$
as follows:~\cite{AdessoG:PRL92:2004,AdessoG:PRA70:2004}
\begin{equation}
  \bs{S}\bs{V}\bs{S}^T=\mathrm{diag}(\nu_{+},\nu_{+},\nu_{-},\nu_{-}),\qquad 
  \nu_{+}\ge\nu_{-}\ge 0,
\end{equation}
where $\nu_{\pm}$ are symplectic eigenvalues. In terms of symplectic
eigenvalues, the physical condition \eqref{eq:phys-cond} can be
expressed as
$$
 \nu_{-}\ge\frac{1}{2}
$$
and the separability condition \eqref{eq:sepa-cond} can be expressed
as
\begin{equation}
 \tilde\nu_{-}\ge\frac{1}{2} \label{eq:sepa-cond2}
\end{equation}
where $\tilde\nu$ represents the symplectic eigenvalue of the
partially transposed covariance matrix $\tilde{\bs{V}}$.  The
logarithmic negativity which measures the degree of the entanglement
is defined by
\begin{equation}
 E_N=-\text{min}\left[\log_2(2\tilde\nu_{-}),0\right].
\end{equation}
If $E_N>0$, the bipartite system is entangled. If $E_N=0$, the
bipartite system is separable.

For the establishment of classicality of the bipartite system, the
separability condition \eqref{eq:sepa-cond2} is necessary but not
sufficient. The separability only means disentanglement of quantum
correlations . For the classicality, the quantum expectation values of
any operators must be reproduced by an appropriate classical
distribution function. Then classical stochastic variables can mimic
the original quantum dynamics. We have discussed in our previous
paper~\cite{NambuY:PRD78:2008} that the condition for the symplectic
eigenvalue
\begin{equation}
  \tilde\nu_{-}\gg \frac{1}{2} \label{eq:classical-condi}
\end{equation}
is required for the system to be regarded as classical. If the system
is separable, there exists a positive normalizable function called the
$P$
function\cite{GardinerCW:S:2004,SimonR:PRL84:2000,DuanL:PRL84:2000}
\begin{align}
  &P(\bs{\xi})=\frac{1}{4\pi^2}\sqrt{\mathrm{det}\bs{P}}
  \exp\left(-\frac{1}{2}\bs{\xi}^T\bs{P}\bs{\xi}\right),\\
  &\quad\bs{P}=\left(\bs{V}+\frac{1}{2}\bs{\Omega}\bs{S}^T\bs{S}\bs{\Omega}^T\right)^{-1},
  \notag
\end{align}
where $\bs{S}\in\mathrm{Sp}(2,\bs{R})\otimes\mathrm{Sp}(2,\bs{R})$ is
the local symplectic transformation of each party and transforms the
covariance matrix $\bs{V}$ to the following standard form
\cite{DuanL:PRL84:2000}
\begin{equation}
  \bs{V}_{II}=\bs{S}\bs{V}\bs{S}^T=
  \begin{pmatrix} a r & & c r & \\
    & a/r & & c'/r \\
    c r & & a r & \\
    & c'/r & & a/r
   \end{pmatrix},\quad
 r=\sqrt{\frac{a-|c'|}{a-|c|}}. 
\end{equation}
Using the $P$ function as a distribution function, it is possible to
calculate the quantum expectation value of the normally ordered
product of any operators
\begin{equation}
  \langle:F(\hat q_A,\hat p_A, \hat q_B, \hat p_B):\rangle
  =\int dq_Adp_Adq_Bdp_BP(q_A,p_A,q_B,p_B)F(q_A,p_A,q_B,p_B).
\end{equation}
If the condition \eqref{eq:classical-condi} is satisfied, the $P$
function acquires the feature of the classical distribution function;
the quantum expectation value for any operators can be reproduced by
using the $P$ function as the distribution function. This implies that
noncommutativity between operators becomes negligible.

\subsection{Entanglement of the quantum field}
\begin{figure}[H]
  \centering
  \includegraphics[width=0.8\linewidth,clip]{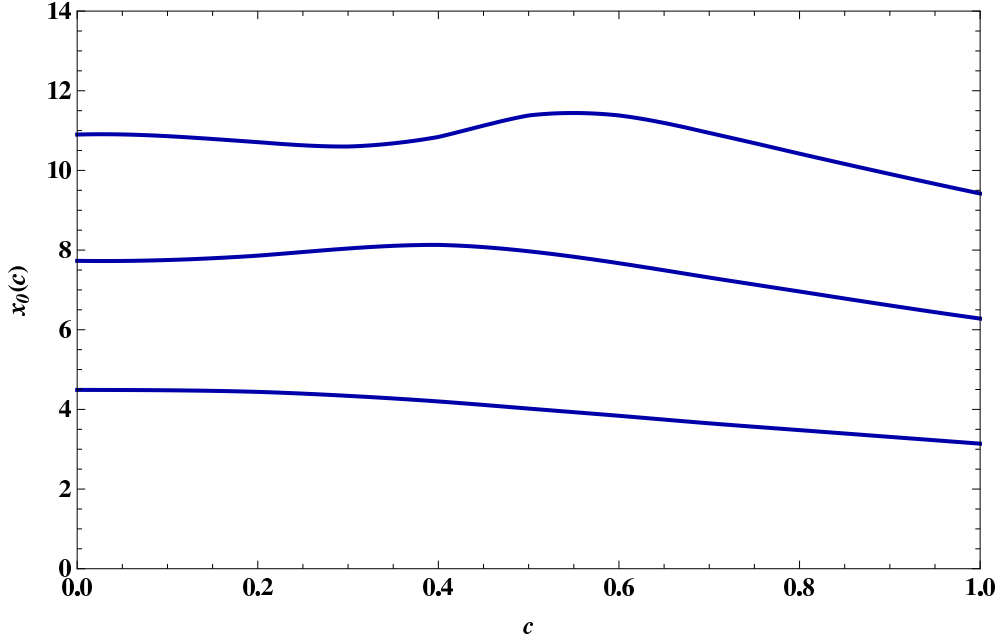}
  \caption{The function $x_0(c)$. Each line corresponds to
    $x_{01}, x_{02}, x_{03}$.}
  \label{fig:x0}
\end{figure}

We consider a massive scalar field in the spatially flat expanding
universe. The metric is
\begin{equation}
 ds^2=-dt^2+a^2(t)d\bs{x}^2=a^2(\eta)(-d\eta^2+d\bs{x}^2).
\end{equation}
The equation of motion for the scalar field is 
\begin{equation}
 \varphi''+\left(m^2a^2-\frac{a''}{a}\right)\varphi-\nabla^2\varphi=0
\end{equation}
where $'$ denotes the derivative with respect to the conformal time
$\eta$.  To define the bipartite system for the scalar field, we
introduce the coarse-grained scalar field using a filter function in
$k$ space. That is, we only include modes with $k_0\le k\le k_c$ in
the Fourier expansion of the scalar field. The lower bound $k_0$ is
the infrared cutoff and corresponds to the system size. The upper
bound $k_c$ is the ultraviolet cutoff and this value determines the
resolution of the measurement. The quantized field $\hat\varphi$ and
its conjugate momentum $\hat p$ can be expressed as
\begin{align}
  &\hat\varphi(\eta,\bs{x})=\int\frac{d^3k}{(2\pi)^{3/2}}W_0\,\theta(k-k_0)
\theta(k_c-k)
\left(
      f_k\hat a_{\bs{k}}+f^*_k\hat
      a^\dag{}_{\!\!\!-\bs{k}}\right)e^{i\bs{k}\cdot\bs{x}}, \label{eq:phi}\\
&\hat p(\eta,\bs{x})=\int\frac{d^3k}{(2\pi)^{3/2}}W_0\,\theta(k-k_0)
\theta(k_c-k)(-i)\left(
      g_k\hat a_{\bs{k}}-g^*_k\hat
      a^\dag{}_{\!\!\!-\bs{k}}\right)e^{i\bs{k}\cdot\bs{x}},\label{eq:pi}\\
& [\hat a_{\bs{k}_1},\hat a^\dag_{\bs{k}_2}]=\del^3(\bs{k}_1-\bs{k}_2),\notag
\end{align}
where $W_0$ is a normalization constant of the filter function. The
mode functions obey
\begin{equation}
  f_k''+\left(k^2+m^2a^2-\frac{a''}{a}\right)f_k=0,\quad
  g_k=i\left(f_k'-\frac{a'}{a}f_k\right),\quad f_kg_k^*+f_k^*g_k=1. 
  \label{eq:modeEq}
\end{equation}
The commutation relation between the coarse-grained fields
\eqref{eq:phi} and \eqref{eq:pi} becomes
\begin{align}
&[\hat\varphi(\eta,\bs{x}_1),
\hat p(\eta,\bs{x}_2)]=  \notag\\
&\qquad 
\frac{iW_0^2}{2\pi^2r^3}\Bigl[
\bigl(\sin(k_c\,r)-(k_c\,r)\cos(k_c\,r)\bigr)
-\bigl(\sin(k_0\,r)-(k_0\,r)\cos(k_0\,r)\bigr)\Bigr], \label{eq:commu}\\
&;\qquad r=|\bs{x}_1-\bs{x}_2|. \notag
\end{align}
For $k_0=0, k_c=\infty,
W_0=1$, the ordinal equal time commutation relation is recovered
\begin{equation}
 [\hat\varphi(\eta,\bs{x}_1), \hat p(\eta,\bs{x}_2)]=i\del^3(\bs{x}_1-\bs{x}_2).
\end{equation}
As our purpose is to define the bipartite system for the quantum
field, we specify two spatial points $\bs{x}_1,\bs{x}_2$ and define
the phase space variables using the scalar field at these points:
\begin{equation}
  \label{eq:variable}
 \hat{\bs{\xi}}=\begin{pmatrix}
   \hat\varphi(\bs{x}_1)\\ \hat p(\bs{x}_1)\\ \hat\varphi(\bs{x}_2)\\
   \hat p(\bs{x}_2)
   \end{pmatrix}.
\end{equation}
The variables $(\hat\varphi(\bs{x}_1), \hat p(\bs{x}_1))$ and
$(\hat\varphi(\bs{x}_2), \hat p(\bs{x}_2))$ correspond to each mode
of the bipartite system.  For these variables to satisfy the condition
of the bipartite system \eqref{eq:2party}, the commutation relation
\eqref{eq:commu} must vanish for $\bs{x}_1\neq\bs{x}_2$ and equals to
be $i$ for $\bs{x}_1=\bs{x}_2$.  The latter condition gives the
normalization of the filter function
$$
 W_0^2=\frac{6\pi^2}{k_c^3-k_0^3}.
$$
To analyze the former condition, we consider the following equation
\begin{equation}
 f_c(x)\equiv\left(\sin x-x\cos x\right)-\left(\sin cx -cx\cos
   cx\right)=0,\quad 0\le c\le 1.
\end{equation}
Let $x_0=x_0(c)$ be the solution of this equation. As shown in
Fig.~\ref{fig:x0}, the function $x_0(c)$ is the multiple valued
function of $c$ and 
\begin{equation}
  x_0=x_{0n}(c),\qquad x_{0n}(1)=n\,\pi,\qquad n=1,2,3,\cdots.
\end{equation}
\noindent
To make the commutation relation vanish at $r\neq 0$, the distance $r$ must
satisfy the following equation
\begin{equation}
 k_cr=x_0\left(\frac{k_0}{k_c}\right),\qquad k_0\le k_c. \label{eq:comm-cond}
\end{equation}
\noindent
Conversely, for any given two points with $r\neq 0$, this relation
provides the scale of the coarse graining $k_0/k_c$ which defines the
bipartite system for the variable~\eqref{eq:variable}. In other words,
if we specify the distance between spatially separated two points, at
which we want to observe the quantum correlation between them, the
scale of the coarse graining of the scalar field is determined by the
relation~\eqref{eq:comm-cond}. If this condition is satisfied, a
measurement of the scalar field as the bipartite system becomes
possible. We rewrite Eq.~\eqref{eq:comm-cond} as
\begin{equation}
  k_0 r=\del\,x_0(\del) \label{eq:comm-cond2}
\end{equation}
where
$$
\del\equiv\frac{k_0}{k_c},\qquad 0\le\del\le 1
$$
determines the scale of the coarse-graining.  For $\del=1~(k_c=k_0)$,
the distance is maximum
\begin{equation}
  k_0 r_{\text{max}}=x_0(1)=n\pi.
\end{equation}
As the value $r_{\text{max}}$ corresponds to the system size related
to the infrared cutoff $k_0$, we must set $n=1$.  Hereafter, we adopt
the smallest branch $x_{01}$ as the function $x_0(c)$.

The correlation functions of the scalar field are given by
\begin{align}
  &c_1\equiv\frac{1}{2}\langle\hat\varphi(\bs{x}_1)\hat\varphi(\bs{x}_2)
  +\hat\varphi(\bs{x}_2)\hat\varphi(\bs{x}_1)\rangle
  =\frac{W_0^2}{2\pi^2}\int_{k_0}^{k_c}\!\!\! 
  dk k^2\left(\frac{\sin kr}{kr}\right)|f_k|^2, \notag \\
  &c_2\equiv\frac{1}{2}\langle
  \hat p(\bs{x}_1)\hat p(\bs{x}_2)+\hat p(\bs{x}_2)\hat p(\bs{x}_1)\rangle
 =\frac{W_0^2}{2\pi^2}\int_{k_0}^{k_c}\!\!\! dk k^2\left(\frac{\sin kr}{kr}\right)|g_k|^2,\\
  &c_3\equiv\frac{1}{2}\langle
  \hat\varphi(\bs{x}_1)\hat p(\bs{x}_2)+\hat p(\bs{x}_2)\hat\varphi(\bs{x}_1)\rangle
  =\frac{W_0^2}{2\pi^2}\int_{k_0}^{k_c}\!\!\! dk k^2
  \left(\frac{\sin kr}{kr}\right)\frac{i}{2}(f_kg_k^*-f_k^*g_k),
  \notag \\
  &a_1=c_1(r=0),\quad a_2=c_2(r=0),\quad a_3=c_3(r=0). \notag
\end{align}
By changing the integral variable to $z=k/k_c$, we have
\begin{align}
  &c_1 =\frac{3}{1-\del^3}\int_{\del}^{1}\!\!\! dz
  z^2j_0(x_0(\del)z)|f_k(\eta)|_{k=k_0z/\del}^2, \notag \\
  &c_2 =\frac{3}{1-\del^3}\int_{\del}^{1}\!\!\! dz
  z^2j_0(x_0(\del)z)|g_k(\eta)|_{k=k_0 z/\del}^2,
  \label{eq:c}\\
  &c_3  =\frac{3}{1-\del^3}\int_{\del}^{1}\!\!\! dz z^2
  j_0(x_0(\del)z)\frac{i}{2}(f_k(\eta)g_k^*(\eta)-f_k^*(\eta)g_k(\eta))|_{k=k_0z/\del}. \notag
\end{align}
They are components of the $4\times 4$ covariance matrix
\eqref{eq:covariance}
$$
 \bs{V}=\begin{pmatrix}\bs{A}&\bs{C}\\
   \bs{C}&\bs{A}\end{pmatrix},\quad
 \bs{A}=\begin{pmatrix} a_1 & a_3 \\ a_3 &a_2\end{pmatrix},\quad
 \bs{C}=\begin{pmatrix} c_1 & c_3 \\ c_3 & c_2 \end{pmatrix}.
$$
Using these components of the covariance matrix $\bs{V}$, the symplectic
eigenvalues are expressed as
\begin{align}
  &(\nu_{-})^2=a_1a_2-a_3^2+c_1c_2-c_3^2-|a_1c_2+a_2c_1-2a_3c_3|,\\
  &(\tilde\nu_{-})^2=a_1a_2-a_3^2-c_1c_2+c_3^2-|(a_1c_2-a_2c_1)^2 \label{eq:symp-eigen}
+4(a_1c_3-a_3c_1)(a_2c_3-a_3c_2)|^{1/2}.
\end{align}

\section{Entanglement of the quantum field in the Minkowski
  spacetime}
As an application of our formalism, we first investigate the
entanglement of the massive scalar field in the Minkowski spacetime.
The mode function for the vacuum state in the Minkowski spacetime is
\begin{equation}
f_k=\frac{1}{\sqrt{2\omega}}\,e^{-i\omega t},\qquad
g_k=i\sqrt{\frac{\omega}{2}}\,e^{-i\omega t},\qquad\omega=\sqrt{k^2+m^2}.
\end{equation}
The correlation functions are
\begin{align*}
  &c_1=\frac{3}{2k_0}\frac{\del}{1-\del^3}\int_\del^1dz
  z^2j_0(x_0(\del)z)\left(z^2+\frac{m^2\del^2}{k_0^2}\right)^{-1/2},\\
  &c_2=\frac{3k_0}{2}\frac{1}{\del(1-\del^3)}\int_\del^1dz
  z^2j_0(x_0(\del)z)\left(z^2+\frac{m^2\del^2}{k_0^2}\right)^{1/2},\\
 &c_3=0.
\end{align*}
The relation between the distance $r$ and the logarithmic negativity
is shown in Fig.~2.
\begin{figure}[H]
  \centering
  \includegraphics[width=1\linewidth,clip]{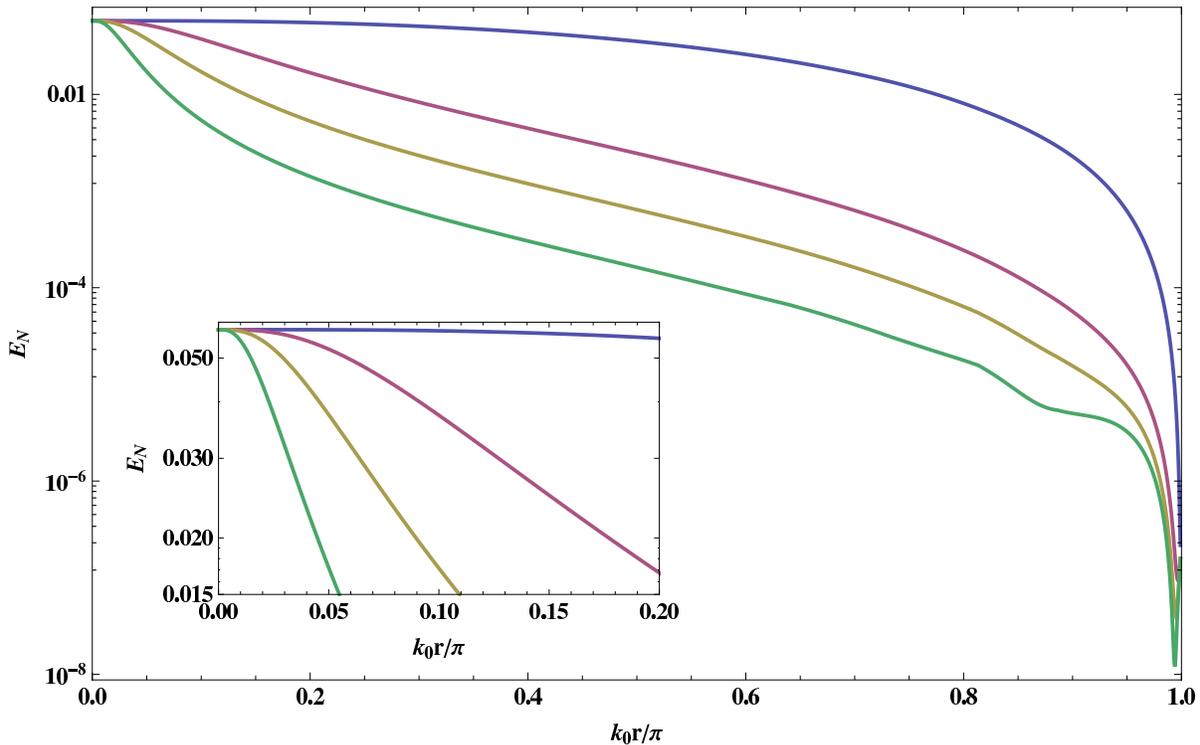}
  \caption{The relation between the spatial distance $r$ and the
    logarithmic negativity $E_N$ [log plot of $E_N(r)$]. Each line
    corresponds to the different value of the mass
    $m/k_0=0~\text{(blue, the upper
      line)},10~\text{(red)},20~\text{(yellow)},40~\text{(green, the
      lower line)} $. The inset is the same plot in the small $r$
    region and shows exponential decay of $E_N$. The decay rate
    depends on the mass $m$ in the small $r$ region.}
\end{figure}
\noindent
For any value of $m$, as the distance increases, the logarithmic
negativity monotonically decreases but does not become zero. This
implies that the Minkowski vacuum is always entangled. For $r\lesssim
r_c\equiv 1/m$, we observe that $r$ dependence of $E_N$ is given by
\begin{equation}
  E_N\sim e^{-r/r_c}
\end{equation}
and the exponential decay rate is proportional to the mass $m$ (see
the inset of Fig.~2).  The entanglement concentrates in the region
with the size of the Compton wavelength $\sim 1/m$.  For $r_c\lesssim
r\ll 1$, the decay law is
\begin{equation}
  E_N\sim e^{-r/r_0}
\end{equation}
where $r_0$ is a constant independent of mass $m$. In this region, the
decay rate of the entanglement is the same for different values of $m$
including the massless case.

\section{Entanglement of the quantum field in the expanding
  universe}
We investigate the effect of the expansion rate of the Universe and
the scalar field mass on the entanglement of the coarse-grained scalar
field.  We assume the following power law expansion of the Universe
\begin{equation}
  a(t)=\left(1+\frac{H_0t}{p}\right)^p,\qquad p>1,\quad H_0>0.
\end{equation}
The conformal time is given by
\begin{equation}
  \eta=\int_0^t\frac{dt}{a}=
  \frac{1}{H_0}\frac{p}{1-p}\left[\left(1+\frac{H_0t}{p}\right)^{1-p}-1\right],
\end{equation}
and in terms of the conformal time, the scale factor is
\begin{equation}
  a(\eta)=\left(\frac{\eta}{\eta_0}+1\right)^{\frac{p}{1-p}},\quad
  \eta_0=\frac{1}{H_0}\frac{p}{1-p}. 
\end{equation}
For the accelerated expansion $p>1$, we have $-\infty<\eta<-\eta_0$
and in the limit of $p\rightarrow\infty$,
$$
 \lim_{p\rightarrow\infty}a=\frac{1}{1-H_0\eta}=\exp(H_0 t).
$$
We set the initial time and the initial scale factor as $t=\eta=0$ and $a_0=1$.

We choose the cutoff parameter for the coarse graining of the scalar
field as follows
\begin{equation}
  k_c=\pi\ep aH=\pi\ep H_0\left(1+\frac{\eta}{\eta_0}\right)^{-1},\qquad
  k_0=\pi H_0. 
\end{equation}
At the initial time $t=\eta=0$, we prepare the spatial region with the size
$H_0^{-1}$ and investigate how the entanglement of the scalar field
between the spatially separated regions evolves as the Universe
expands. We introduce the parameter $\ep$ to specify the scale of
coarse graining and this parametrization is conventionally used for
the stochastic approach to inflation~\cite{StarobinskiA:1986}. In our
analysis, this parameter must satisfy
\begin{equation}
  \frac{H_0}{a\,H}\le \ep
\end{equation}
which comes from $k_0\le k_c$. The value $\ep $ need not be smaller
than unity that is usually assumed for the stochastic approach to
inflation.  We calculate the symplectic eigenvalue $\tilde\nu_{-}$ as
a function of the physical distance
\begin{equation}
r_{\text{phys}}=ar=\frac{1}{\pi\ep H}x_0\left(\frac{H_0}{\ep a H}\right)
\end{equation}
and the e-folding $N=\ln(a/a_0)$. What we are interested in is the
condition of the separability~(\ref{eq:sepa-cond2}) and the
classicality~(\ref{eq:classical-condi}). We plot these conditions in
the $(r_{\text{phys}}, N)$ space.

\subsection{The effect of expansion rate on the entanglement}
We first investigate the effect of the expansion rate of the Universe
on the entanglement of the massless scalar field. In our previous
paper~\cite{NambuY:PRD78:2008}, we used a lattice model of the
massless scalar field and found that the bipartite system becomes
separable when the size of the spatial region exceeds the Hubble
horizon. We aim to confirm this behavior for the accelerated universe
with the power law expansion. The mode equation for the massless field
is
\begin{align}
 &f_k''+\left(k^2-\frac{\al^2-1/4}{(\eta+\eta_0)^2}\right)f_k=0,\qquad
\al^2=\frac{1}{4}\left(\frac{3p-1}{p-1}\right)^2,\\
 &g_k=i\left(f_k'-\frac{p}{1-p}\frac{f_k}{\eta+\eta_0}\right) \notag
\end{align}
As the quantum state of the scalar field, we choose the Bunch-Davis
vacuum state, the mode function is given by
\begin{align}
 &f_k=\frac{\sqrt{\pi}}{2}e^{i(2\al+1)\pi/4}(-(\eta+\eta_0))^{1/2
}H_{\al}^{(1)}(-k(\eta+\eta_0))
,\\
 &g_k=-i\frac{\sqrt{\pi}}{2}e^{i(2\al+1)\pi/4}k(-(\eta+\eta_0))^{1/2}
H_{\al-1}^{(1)}(-k(\eta+\eta_0)). 
\end{align}

\begin{figure}[H]
  \centering
  \includegraphics[width=0.9\linewidth,clip]{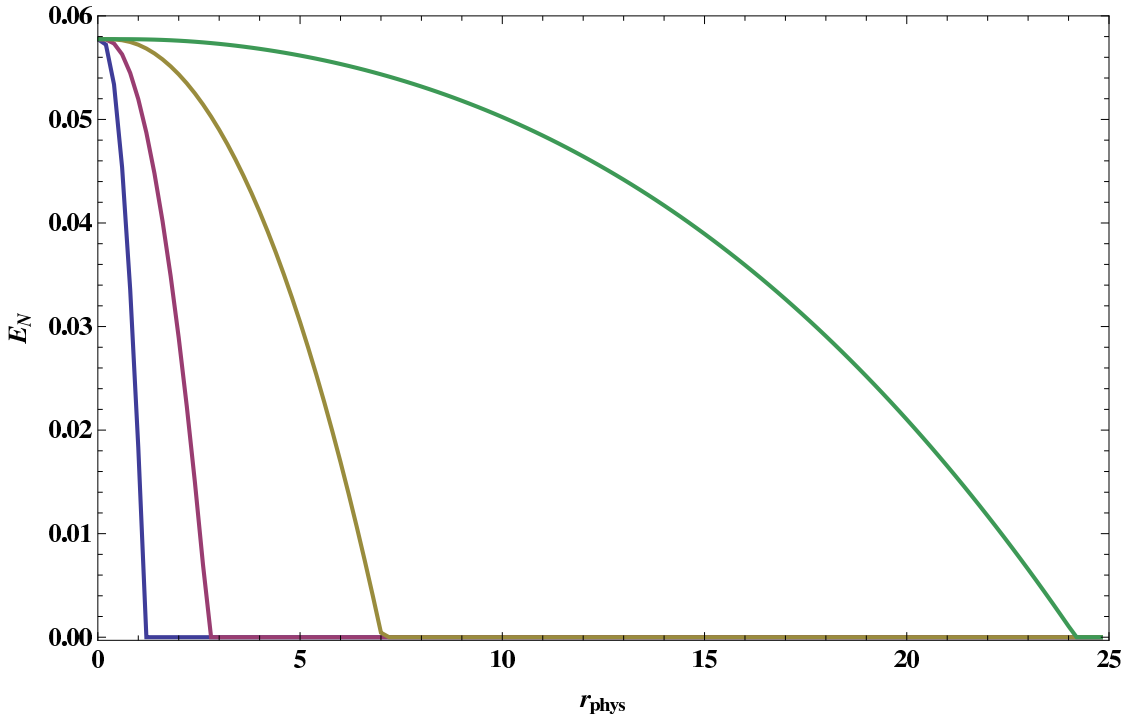}
  \caption{Spatial dependence of the logarithmic negativity at $N=10$
    for the massless field. The distance $r_{\text{phys}}$ is in the
    unit of $H_0^{-1}$. $E_N$ goes to zero at
    $r_{\text{phys}}=r_{\text{separable}}$. Each line corresponds to
    $p=100~\text{(blue, the lower line)}, 10~\text{(red)}, 5~\text{(yellow)},
    3~\text{(green, the upper line)}$.}
  \label{fig:nuRpow}
\end{figure}

We first present the spatial dependence of the logarithmic negativity
$E_N$ at the e-folding $N=10$ for the power index $p=100,10,5,3$
(Fig.~\ref{fig:nuRpow}). $E_N$ decays as $r_{\text{phys}}$ increases
and becomes zero at $r_{\text{phys}}=r_{\text{separable}}$. For large
spatial separation $r_{\text{separable}}<r_{\text{phys}}$, $E_N=0$ and
the system is separable.  We numerically check that the $p$ dependence
of $r_{\text{separable}}$ is given by
\begin{equation}
  r_{\text{separable}}\approx H_0^{-1}\exp\left(\frac{N}{p}\right)=H^{-1}.
\end{equation}

\noindent
Hence, the horizon scale gives the scale of the separability and this
result is consistent with our previous analysis using a lattice
model~\cite{NambuY:PRD78:2008}.

\begin{figure}[H]
  \centering
  \includegraphics[width=0.49\linewidth,clip]{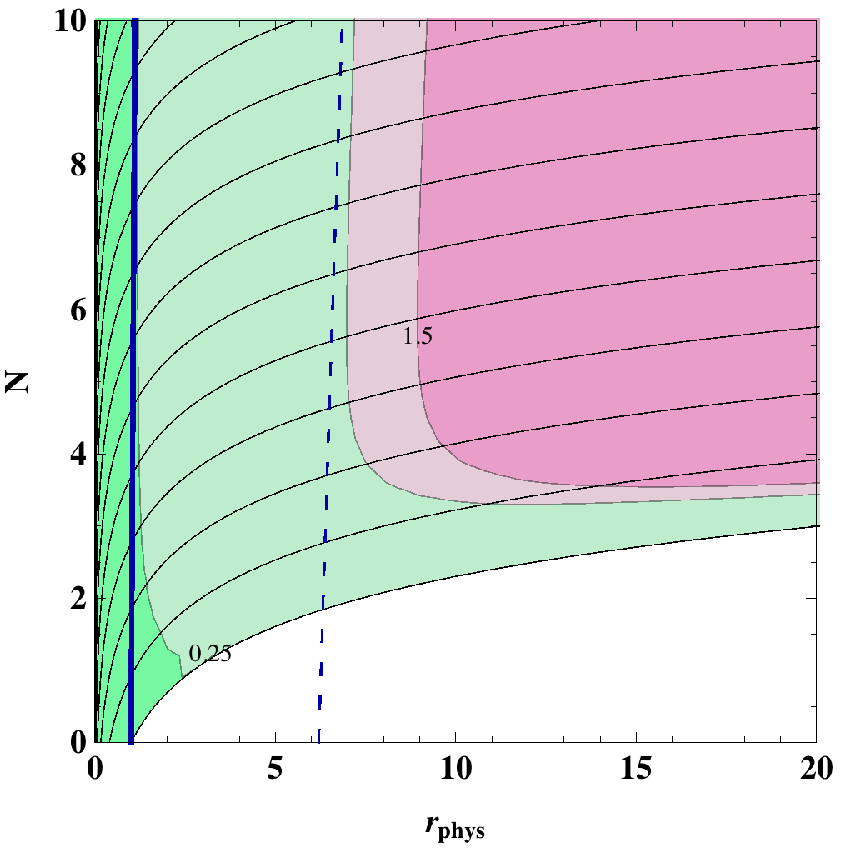}
  \includegraphics[width=0.49\linewidth,clip]{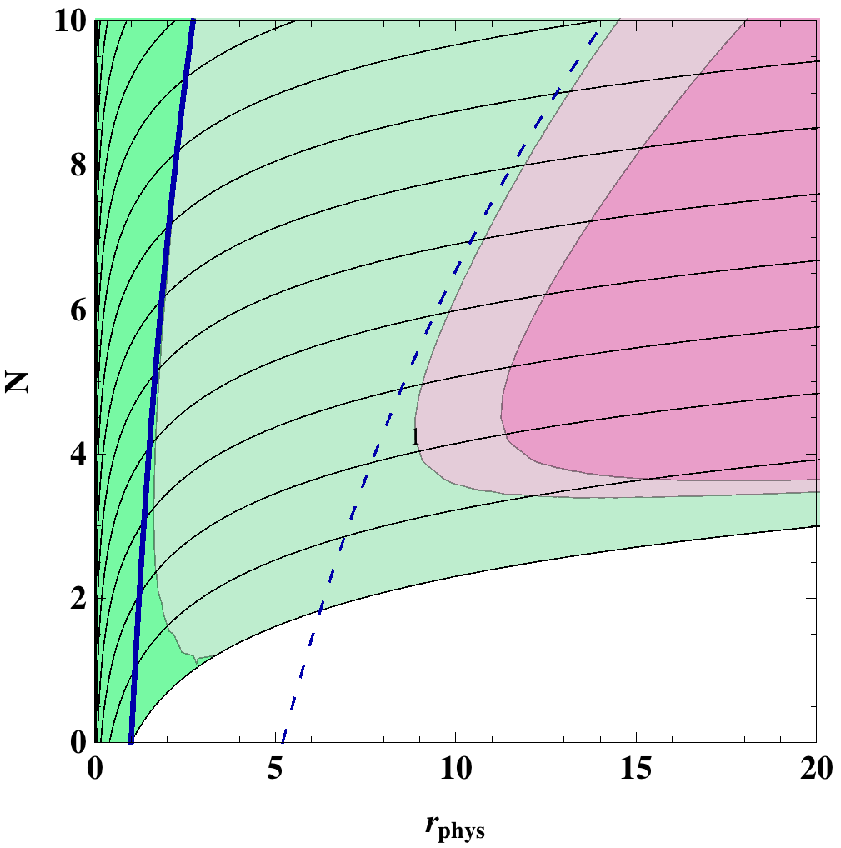}
  \includegraphics[width=0.49\linewidth,clip]{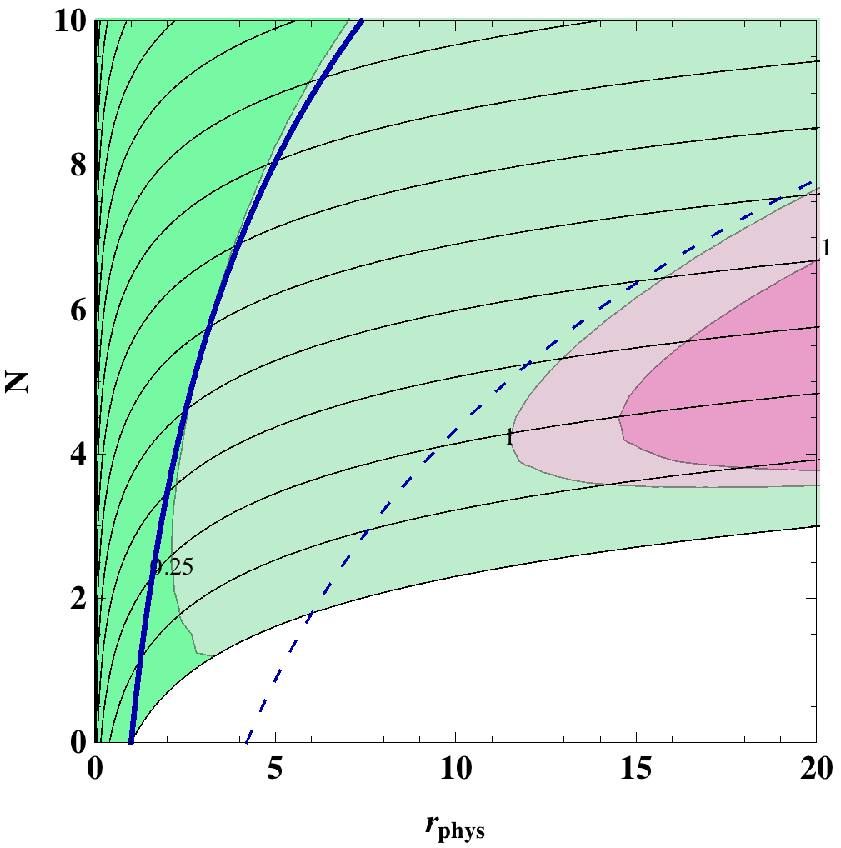}
  \includegraphics[width=0.49\linewidth,clip]{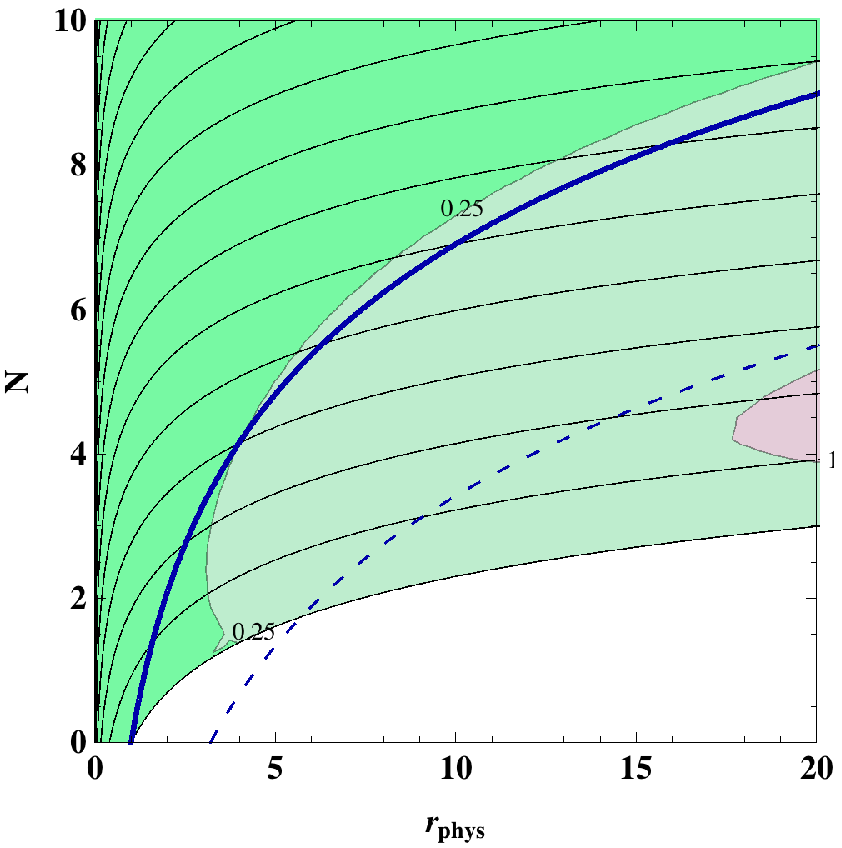}
  \caption{The value of the symplectic eigenvalue $(\tilde\nu_{-})^2$
    in the $(r_{\text{phys}},N)$ space for the massless scalar field
    in the Universe with power law expansion. The distance
    $r_{\text{phys}}$ is in the unit of $H_0^{-1}$. Each panel
    corresponds to the different expansion rate $p=100, 10, 5, 3$ from
    the top left to the down right. The dark green region corresponds
    to $(\tilde\nu_{-})^2<1/4$ and the system is entangled in this
    region. The light green region corresponds to
    $1/4<(\tilde\nu_{-})^2<1$ and the system is separable but the
    classicality condition is not satisfied. The pink region
    corresponds to $1<(\tilde\nu_{-})^2$. The blue solid line is
    $r_{\text{phys}}=H^{-1}$ (horizon scale) and the black solid lines
    represent different comoving scales.}
  \label{fig:nuP}
\end{figure}

In Fig.~\ref{fig:nuP}, we show the behavior of the symplectic
eigenvalue $(\tilde\nu_{-})^2$ in the $(r_{\text{phys}},N)$ space for
the power index $p=100,10,5,3$. For the Universe with the power law
expansion, the horizon scale $H^{-1}$ changes with time.  We observe
that the line of the separability condition $ (\tilde\nu_{-})^2=1/4$
asymptotically coincides with the horizon line (the blue solid
line). When the distance between two points is smaller than the
horizon $H^{-1}$, they are entangled and they become disentangled
after their separation exceeds the horion length.
\noindent
This behavior of disentanglement does not depend on the expansion
rate $p$ and the condition of the separability is determined by
$r_{\text{phys}}\approx H^{-1}$.  Thus, for any value of $p>1$, for
sufficiently large value of e-folding, the boundary between
separable and entangled region coincides with the horizon line
$H^{-1}$. This means the accelerated expansion of the Universe or the
existence of the horizon determines the property of the separability
of the massless scalar field in the expanding universe.  

The line $(\tilde\nu_{-})^2=1$ giving the criterion of classicality
asymptotically approaches several times larger than the horizon
scale. In the region $(\tilde\nu_{-})^2>1$, the noncommutativity
between canonical variables can be neglected when we evaluate the
expectation values of operators and be consistent with the result
obtained in the previous analysis for the behavior of the each
comoving wave
mode\cite{GuthAH:PRD32:1985,AlbrechtA:PRD50:1994,PolarskiD:CQG13:1996,LesgourguesJ:NPB497:1997,KieferC:CQG15:1998};
for superhorizon scale quantum fluctuations, the noncommutativity
between canonical variables becomes negligible because the growing
mode solution is dominant and we can neglect $\hbar$ in the
uncertainty relation. We confirmed the equivalent condition for the
classicality from the condition of the existence of the classical
distribution function and the symplectic eigenvalues.

\subsection{The effect of the mass on the entanglement}
We next investigate the effect of the mass of the scalar field on the
entanglement.  For the massive scalar field in the de Sitter spacetime
($p=\infty$), the mode equation becomes
\begin{align}
 &f_k''+\left(k^2-\frac{\al^2-1/4}{(\eta+\eta_0)^2}\right)f_k=0,\qquad
\al^2=\frac{9}{4}-\frac{m^2}{H_0^2},\\
 &g_k=i\left(f_k'+\frac{f_k}{\eta+\eta_0}\right),\qquad\eta_0=-\frac{1}{H_0} \notag
\end{align}
Assuming the Bunch-Davis vacuum state, the mode function is given by
\begin{align}
 &f_k=\frac{\sqrt{\pi}}{2}e^{i(2\al+1)\pi/4}(-(\eta+\eta_0))^{1/2}
H_{\al}^{(1)}(-k(\eta+\eta_0))
,\\
 &g_k=i\frac{\sqrt{\pi}}{2}e^{i(2\al+1)\pi/4}(-(\eta+\eta_0))^{-1/2}\\
 &\qquad \times\left[\left(\al
-\frac{3}{2}\right)H_{\al}^{(1)}(-k(\eta+\eta_0))+k(\eta+\eta_0)
H_{\al-1}^{(1)}(-k(\eta+\eta_0))\right]. \notag
\end{align}

\begin{figure}[H]
  \centering
  \includegraphics[width=0.9\linewidth,clip]{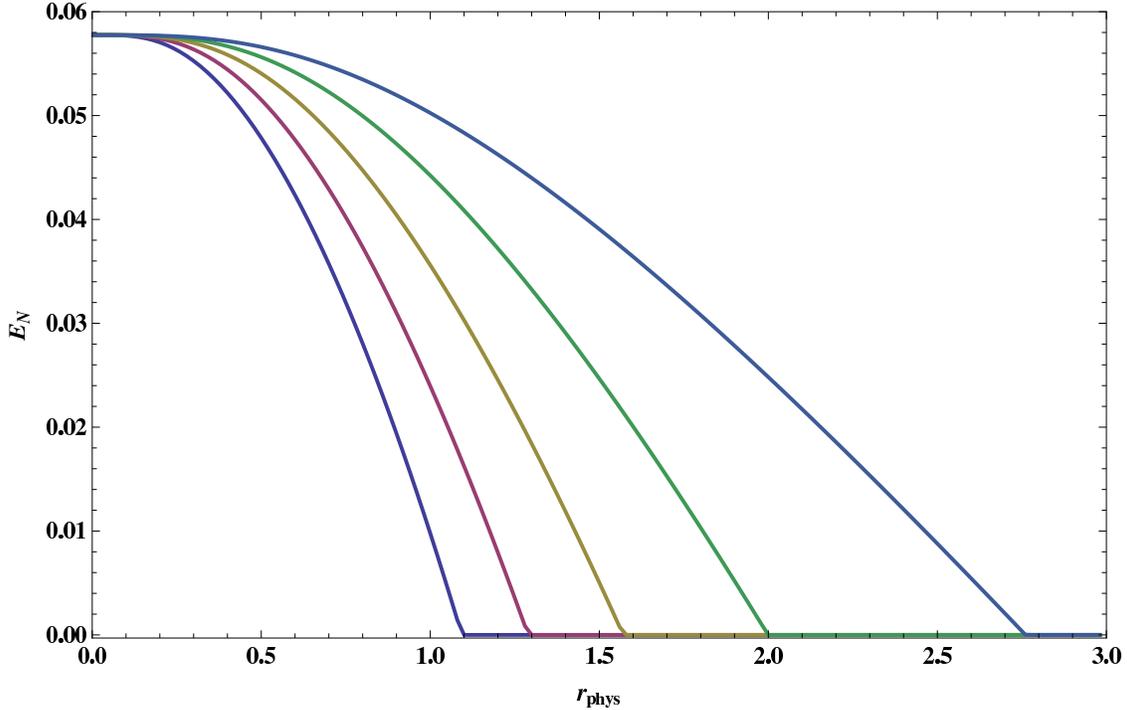}
  \caption{Spatial dependence of the logarithmic negativity at $N=10$
    for the massive scalar field in the de Sitter spacetime. The
    distance $r_{\text{phys}}$ is in the unit of $H_0^{-1}$. Each line
    corresponds to $m^2/H_0^2=0~\text{(blue, the lower line)}, 1/4~\text{(red)},
    1/2~\text{(yellow)}, 3/4~\text{(green)}, 1~\text{(blue, the upper line)}$.}
  \label{fig:enRm}
\end{figure}

Fig.~\ref{fig:enRm} shows the spatial dependence of $E_N$ at $N=10$
for $m^2/H_0^2=0,1/4,1/2,1$. $E_N$ decays as $r_{\text{phys}}$
increases and becomes zero at
$r_{\text{phys}}=r_{\text{separable}}$. For large spatial separation
$r_{\text{separable}}<r_{\text{phys}}$, $E_N=0$. We observe that the
mass dependence of $r_{\text{separable}}$ is given by
\begin{equation}
  r_{\text{separable}}\approx
  H_0^{-1}\left(1-c_0\,\frac{m^2}{2H^2}\right)^{-1/2},\qquad c_0\sim
  1.4. \label{eq:rsepM}
\end{equation}
\noindent
For $m\neq 0$, $r_{\text{separable}}$ does not coincide with the
horizon scale $H_0^{-1}$, which is the characteristic scale of the
disentanglement for the massless scalar field.  The mass
dependence~(\ref{eq:rsepM}) of $r_{\text{separable}}$ can be
understood as follows. Let us recall the form of the mode
equation~(\ref{eq:modeEq}). The mode changes its behavior depending on
the following wave numbers:
\begin{equation}
  \left(\frac{k_*}{a}\right)^2\equiv \frac{a''}{a^3}-m^2.
\end{equation}
For $k>k_*$, the mode behaves oscillatory and for $k<k_*$, the mode
becomes unstable and frozen. This critical wave number corresponds to
the physical length
\begin{equation}
  r_*=\frac{1}{H_0}\left(1-\frac{m^2}{2H_0^2}\right)^{-1/2}.
\end{equation}
If the physical wavelength of the scalar field is smaller than this
length, the scalar field behaves oscillatory and then becomes frozen
after its wavelength exceeds $r_*$ by the cosmic expansion. For the
massless case, $r_*$ coincides with the horizon length $H^{-1}$ and
the nonzero mass increases the length $r_*$. Our numerical
result~(\ref{eq:rsepM}) indicates
\begin{equation}
  r_{\text{separable}}\approx r_*.
  \label{eq:rsepM2}
\end{equation}

\begin{figure}[H]
  \centering
  \includegraphics[width=0.48\linewidth,clip]{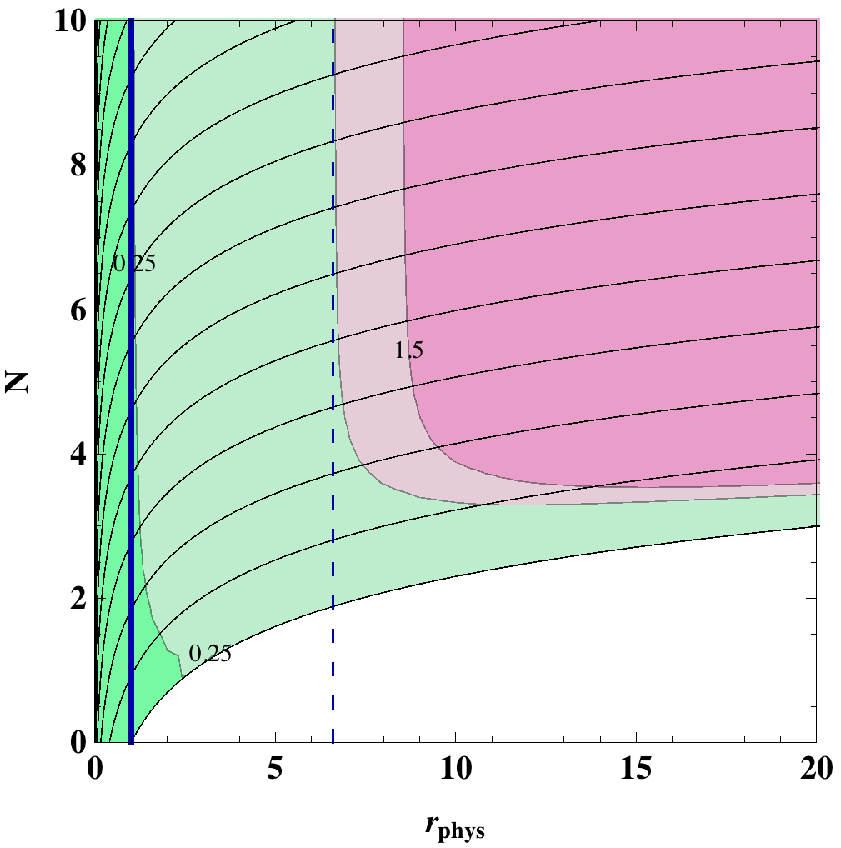}
  \includegraphics[width=0.48\linewidth,clip]{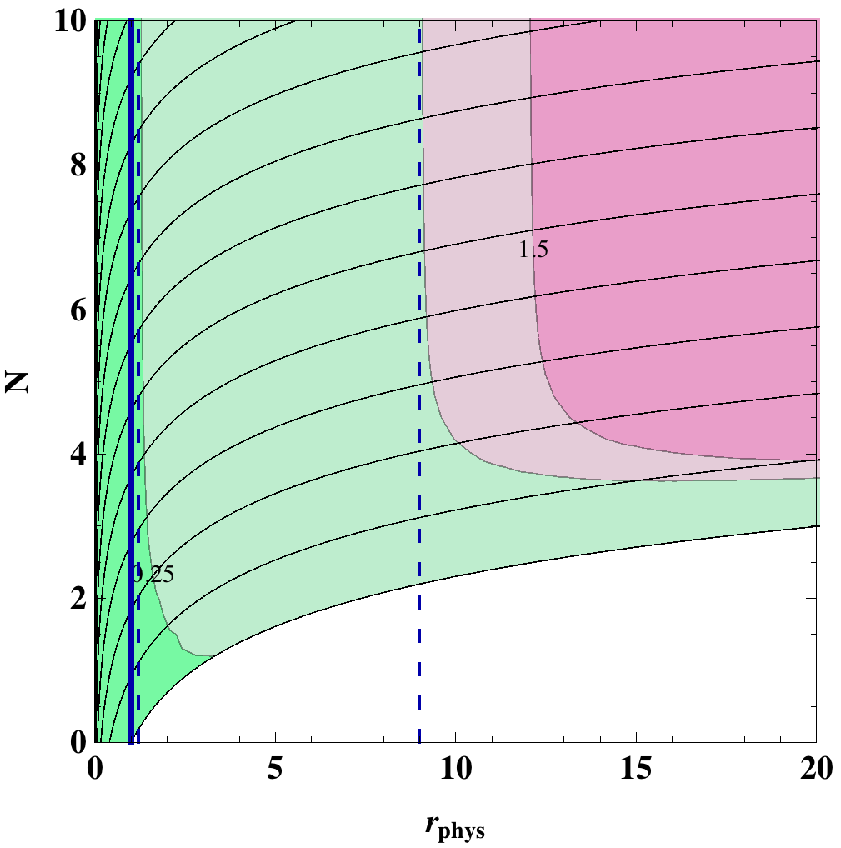}
  \includegraphics[width=0.48\linewidth,clip]{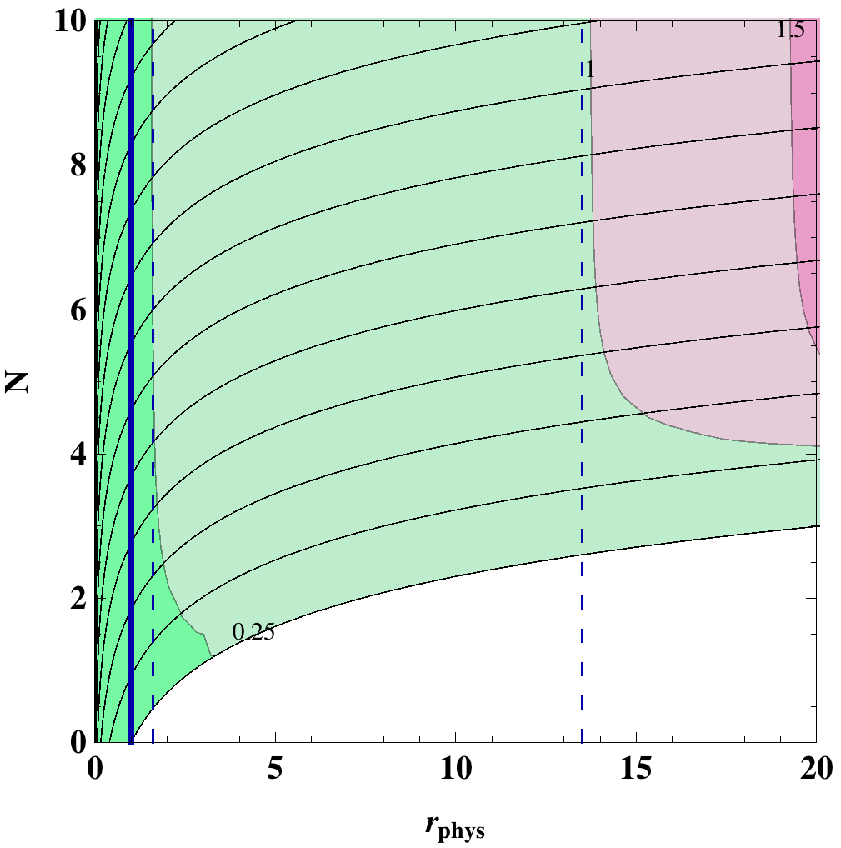}
  \includegraphics[width=0.48\linewidth,clip]{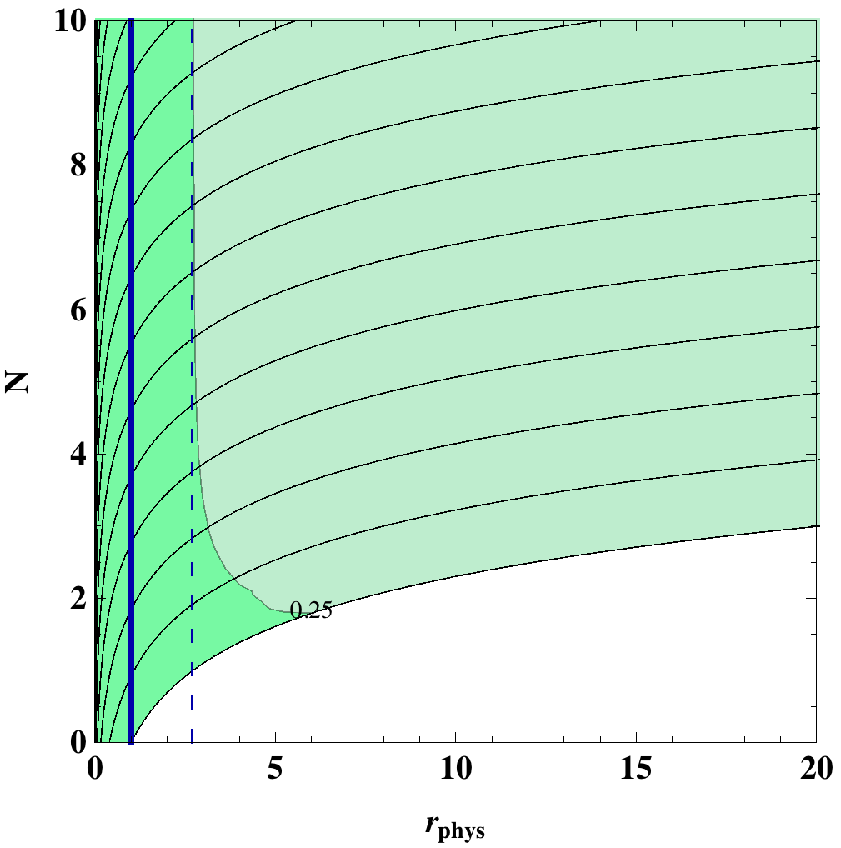}
  \caption{The $(r_{\text{phys}},N)$ dependence of the symplectic
    eigenvalue $(\tilde\nu_{-})^2$ for the massive scalar field in the
    de Sitter spacetime. The distance $r_{\text{phys}}$ is in  units
    of $H_0^{-1}$. Each panel show $(\tilde\nu_{-})^2$ for
    $m^2/H_0^2=0,1/4,1/2,1$ from the top left to the down right. The
    dark green region corresponds to $(\tilde\nu_{-})^2<1/4$ and the
    system is entangled. The light green region corresponds to
    $(\tilde\nu_{-})^2>1/4$ and the system is separable. The pink
    region satisfies the condition of the classicality
    $(\tilde\nu_{-})^2>1$. The black solid lines represent the
    different comoving scales. The line $(\tilde\nu_{-})^2=1/4$
    deviates from the horizon scale $H_0^{-1}$ (the blue solid line)
    as the mass increases.}
  \label{fig:nuM}
\end{figure}

Figure~\ref{fig:nuM} shows the $(r_{\text{phys}},N)$ dependence of the
symplectic eigenvalue $(\tilde\nu_{-})^2$.  For the massless case, for
the sufficiently large value of the e-folding, the system becomes
separable after the physical distance between two points exceeds the
horizon $H_0^{-1}$.  As the mass increases, the line
$(\tilde\nu_{-})^2=1/4$ representing the separability condition
deviates from the horizon line $H_0^{-1}$ as expected from
\eqref{eq:rsepM2}. The line $\tilde\nu_{-}=1$, which gives the criterion
of the classicality~\eqref{eq:classical-condi}, corresponds to the
scale $6\sim 7$ times larger than the horizon size.

We compare this behavior of the mass dependence on the entanglement
with the Minkowski case. For the Minkowski spacetime, the
characteristic size of the entangled region is given by the Compton
wavelength $1/m$ and this size decreases as the mass increases. For
the de Sitter case, if we consider a sufficiently small region
compared to the horizon length, we can neglect the effect of the
cosmic expansion and the behavior of the entanglement is the same as the
Minkowski case. For larger scales $r_*<r_{\text{phys}}$, the system
becomes separable and this disentanglement behavior does not occur in
the Minkowski spacetime. The size of the entangled region is larger
than the horizon scale and increases as the mass increases.  We expect
that this behavior of the entanglement is related to the causal structure
of the de Sitter spacetime.

\subsection{Classicality condition and scale of coarse graining}
In this subsection, we discuss the relation between the classicality
and the scale of coarse graining. As we have already observed, the
coarse graining with a sufficiently large scale leads to
$\tilde\nu_{-}\gg 1/4$ and the system becomes classical. To investigate
the effect of the scale of coarse-graining  on the classicality, we
plot the symplectic eigenvalues as the function of $\del=k_0/k_c$ in
Fig.~\ref{fig:eigen-comov}:
\begin{figure}[H]
  \centering
  \includegraphics[width=0.48\linewidth,clip]{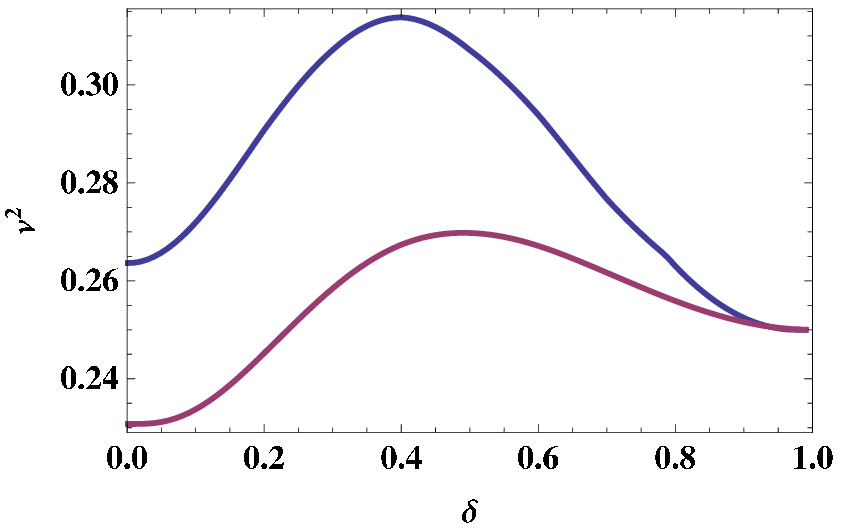}
  \includegraphics[width=0.48\linewidth,clip]{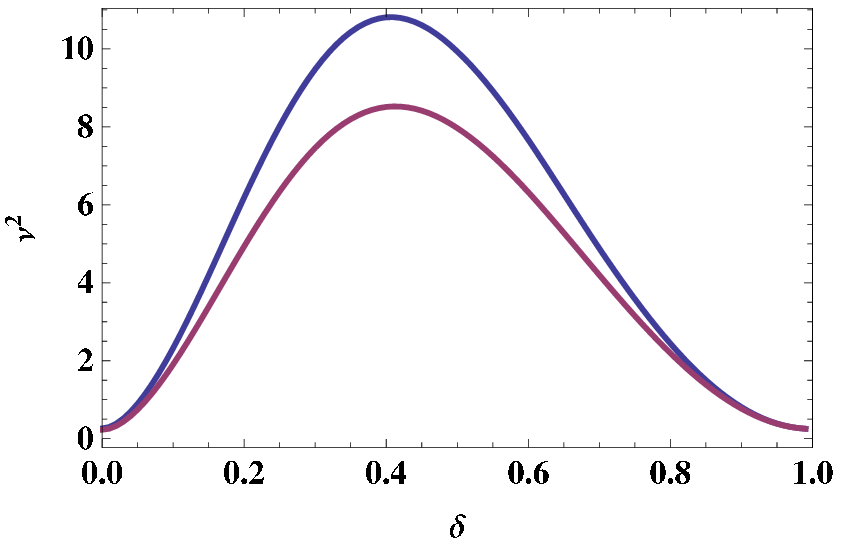}
  \caption{The dependence of the comoving scale $\del$ of the
    symplectic eigenvalues for the massive scalar field in the de
    Sitter universe. The left panel shows $(\nu_{-})^2$ (blue line)
    and $(\tilde\nu_{-})^2$ (red line) at $N=2$. The right panel is at
    $N=5$. The mass of the scalar field is $m^2/H_0^2=1/4$. At any
    time, $\tilde\nu_{-}>1/2$ for $1/a<\del\le 1$.}
  \label{fig:eigen-comov}
\end{figure}
\noindent
The relation between $\del$ and the coarse-graining parameter $\ep$ is
\begin{equation}
  \del=\frac{H_0}{\ep\,a\,H}\le 1.
\end{equation}
As the physical distance is
\begin{equation}
  ar=\frac{1}{\pi\ep H}x_0\left(\frac{H_0}{\ep a H}\right)\sim\frac{1}{\ep H},
\end{equation}
$\del$ represents the comoving scale of the bipartite system.  For the
super horizon scale $H_0/(aH)<\del\le 1$, as we have already
confirmed, the separability condition $(\tilde\nu)^2>1/4$ is
satisfied. For late time $a\gg 1$, the classicality condition
$(\tilde\nu)^2\gg 1/4$ is also satisfied for a wide range of the
comoving scale $\del$. However, as is shown in the right panel of
Fig.~\ref{fig:eigen-comov}, for too large a value of $\del$, the
classicality condition is not satisfied as $\nu,\tilde\nu$ are
decreasing functions of $\del$ for $\del\sim 1$.  Thus, we have the
maximum scale of the coarse graining to retain the classicality. We can
estimate this scale using the asymptotic form of the correlation
functions ~\eqref{eq:c} and the definition of the symplectic
eigenvalue~(\ref{eq:symp-eigen}). Assuming the large scale
coarse graining $\ep\ll 1$, we obtain the following asymptotic form of
the symplectic eigenvalue $\nu$ and $\tilde\nu$ for the massive scalar
field in the de Sitter spacetime:
\begin{equation}
  \nu^2, \tilde\nu^2\sim
  \begin{cases}
    & (a\,\del)^{4\al-4}=\left(\dfrac{1}{\ep}\right)^{2}
\qquad (\text{ for}~\del\sim 0),\\
    & a^{4\al-4}(1-\del)^2+1/4 =\ep^{4m^2/(3H_0^2)}\left(a-\dfrac{1}{\ep}\right)^2+1/4\qquad
    (\text{ for}~\del\sim 1)
  \end{cases}
\end{equation}
where we have used $\al\approx 3/2-m^2/(3H_0^2), ~m^2/H_0^2\ll 1$. 
For the small comoving scale $\del\sim 0$, the classicality condition
$\tilde\nu^2\gg 1/4$ requires $\ep\ll 1$ and this is consistent with
$\del\sim 0$ provided that $a\gg 1$ (late time). Thus, sufficiently large scale
coarse graining $\ep\ll 1$ is necessary to obtain the classicality of the scalar
field. For the large comoving scale $\del\sim 1$, to keep
$\tilde\nu^2\gg 1/4$,
\begin{equation}
  1\ll\ep^{2m^2/(3H_0^2)}=e^{2m^2/(3H_0^2)\ln\ep}
\end{equation}
is necessary and this yields the lower bound of $\ep$:
\begin{equation}
  e^{-3H_0^2/(2m^2)}\ll\ep.
\end{equation}
Therefore, we need the following condition for the coarse-graining
parameter to guarantee the classicality of the coarse-grained field
\begin{equation}
  e^{-3H_0^2/(2m^2)}\ll\ep\ll 1.   \label{eq:classical-cond1}
\end{equation}
For the massless scalar field in the Universe with a power law
expansion, we have 
\begin{equation}
  \nu^2, \tilde\nu^2\sim
  \begin{cases}
    & \left(\dfrac{a\,\del\, H}{H_0}\right)^{4\al-4}=\left(\dfrac{1}{\ep}\right)^{2}
\qquad (\text{ for}~\del\sim 0),\\
    & \left(\dfrac{aH}{H_0}\right)^{4\al-4}(1-\del)^2+1/4 
    =\ep^{-4/p}\left(\dfrac{aH}{H_0}-\dfrac{1}{\ep}\right)^2+1/4\qquad
    (\text{ for}~\del\sim 1)
  \end{cases}
\end{equation}
where we have used $\al\approx 3/2+1/p, ~p\gg 1$.  The condition
$\tilde\nu^2\gg 1/4$ leads to
\begin{equation}
  \label{eq:classical-cond2}
  \ep\ll 1.
\end{equation}
The condition \eqref{eq:classical-cond1} and
\eqref{eq:classical-cond2} for the coarse-graining parameter $\ep$ are
the same as the ones that appeared in the stochastic approach to
inflation~\cite{StarobinskiA:1986,SasakiM:NPB308:1988,HabibS:PRD46:1992}
to ensure the amplitude of the stochastic noise is independent of the
coarse-graining parameter. With these conditions, the stochastic
calculus based on the Langevin equation reproduces the field theoretic
result of expectation values. However, in the context of the
stochastic approach, it was not clear why these conditions guarantee
the validity of the stochastic approach. From the view point of the
entanglement and the classicality of the quantum field, the conditions 
\eqref{eq:classical-cond1} and \eqref{eq:classical-cond2} are
equivalent to $\tilde\nu^2\gg 1/4$; with this condition, the
coarse-grained quantum field becomes separable and there exists a
classical distribution function which reproduces the expectation
values of the original quantum system. In other words, we have
appropriate classical stochastic variables or stochastic processes
that mimic the original quantum dynamics. This supports the validity
of the stochastic approach which treats the quantum field as the
classical stochastic variables.


\section{Summary and conclusion}

We investigated the behavior of the bipartite entanglement of the
scalar field in the expanding universe.  To define the bipartite
system for the quantum field, we introduced the coarse graining of the
scalar field. In our formalism, the scale of the coarse graining
corresponds to the spatial distance between two points at which we
want to measure the bipartite entanglement. This defines the bipartite
system with the two mode Gaussian state and we can judge the
separability of the system by the criterion based on the partial
transpose operation.

For the massless field, the disentanglement occurs when the scale of
the coarse graining equals to the horizon length $H^{-1}$. The horizon
scale determines the causal structure of the accelerated expanding
universe and two points are causally disconnected beyond this
scale. We have confirmed that the quantum correlation or the bipartite
entanglement disappears beyond this scale for the massless scalar
field. This disentanglement behavior is necessary for the quantum
field to acquire the classical nature.  With inclusion of the mass of
the scalar field, we found that the mass increases the scale of the
disentanglement.  The system becomes separable when the oscillatory
behavior of the mode function stops and changes to be frozen. This
scale is larger than the horizon length and corresponds to the sonic
horizon which discriminates the behavior of the mode function.  After
the disentanglement occurs and the system becomes separable, the
classicality condition is satisfied at a sufficiently late time or for
sufficiently large scale coarse graining. We derived the condition for
the scale of the coarse graining  needed to satisfy the
classicality condition at late time and found that the upper and the
lower bound for the coarse-graining parameter.  These bounds are
equivalent to ones that appeared in the stochastic approach to inflation to
guarantee the cut-off independence of the stochastic dynamics of the
scalar field.

After the classicality condition is satisfied, it is possible to
calculate the quantum expectation value of any operators using the
classical distribution functions such as the $P$ function and the
Wigner function. However, this does not mean that the information on
the quantum correlation or the entanglement before the
classicalization is lost. The remnant of the quantum correlation is
encoded in the classical distribution function and this is responsible
for the origin of structure in our Universe.  It will be interesting
to investigate the relation between the classical stochastic property
of the fluctuation after the classicalization and the encoded quantum
correlation. The analysis towards such a direction will make clear the
mechanism of the quantum to classical transition of the quantum
fluctuation in the inflationary universe.

\begin{acknowledgments}
  This work was supported in part by the JSPS Grant-In-Aid for
  Scientific Research [C] (19540279) and the Grant-in-Aid for Nagoya
  University Global COE Program, ``Quest for Fundamental Principles in
  the Universe: from Particles to the Solar System and the Cosmos,''
  from the Ministry of Education, Culture, Sports Science and
  Technology of Japan.
\end{acknowledgments}


\end{document}